\documentclass{ws-tpe}
\usepackage{multicol}

\newcommand{\specialcell}[2][c]{%
	\begin{tabular}[#1]{@{}c@{}}#2\end{tabular}}

\begin{document}

\catchline{1}{1}{2019}{}{}
\markboth{M. S. Feser \& D. Höttecke}{Development of a test in German language to assess middle school students’ physics proficiency}

\title{Development of a test in German language to assess middle school students’ physics proficiency}

\author{Markus Sebastian Feser}
\address{Physics Education, Universität Hamburg\\
	Hamburg, Germany\\
	\email{markus.sebastian.feser@uni-hamburg.de}}

\author{Dietmar Höttecke}
\address{Physics Education, Universität Hamburg\\
	Hamburg, Germany\\
	\email{dietmar.hoettecke@uni-hamburg.de}}

\maketitle


\begin{abstract}
This short contribution reports the development of a test for assessing middle school students’ physics proficiency via multiple-choice single-select items in German language. The test assesses students’ content and procedural knowledge across various content areas that are typical of physics education at the middle-school level and is based on adapted items developed within the Third International Mathematics and Science Study (TIMSS). We report the study design we used to develop this test, as well as the results and selected parameters regarding the test’s psychometric quality.
\end{abstract}

\keywords{secondary physics education; test development; physics proficiency; TIMSS}

\begin{multicols}{2}
\section{Theoretical background and aim of the study}
Students’ learning in the physics classroom strongly depends on their physics proficiency, i.e., the entirety of their previously acquired content and procedural knowledge in physics.\cite{1,2,3} As a result, one of the standard procedures of empirical research on physics education is to assess students’ physics proficiency via achievement tests in order to use this information, e.g., as a covariate. However, unlike in other school subjects, there is a lack of subject-specific achievement tests for physics in German language that are generally accessible in educational research.\cite{4,5} To fill this gap, a popular approach among physics education researchers in Germany (e.g., Ref.~\refcite{6,7,8,9,10}) is to assess students’ proficiency using self-compiled pencil-and-paper tests based on published items from the Third International Mathematics and Science Study (TIMSS) (for details on the science achievement construct behind TIMSS, see Ref.~\refcite{11}). Unfortunately, only a small number of TIMSS items have been published in German language.\cite{12} Therefore, based on the released TIMSS item sets, it is hardly possible to design physics achievement tests in German language that are of sufficient psychometric quality. In addition to multiple-choice single-select items, TIMSS also uses open-ended items. These open-ended items require a comparatively long time on task on the part of students. Moreover, the data analysis of open-ended items takes time and effort because the data must be coded carefully by several researchers.\cite{13,14} Therefore, items from the pool of published TIMSS items that are open-ended are less suitable for assessing students’ physics proficiency in a highly test-economic and objective manner as compared to multiple-choice items.

In this short contribution, we report the results of a research project that was aimed at tackling these issues. More precisely, we report on the development of a test in German language for assessing middle school students’ physics proficiency. This test features the following characteristics:
\begin{romanlist}[(ii)]
	\item The test consists of 20 items in total and has an estimated time requirement of 20 minutes.
	\item The test is based on adapted items developed for the assessment of primary- and middle-school students’ science achievement within TIMSS.
	\item To ensure test economy and objectivity, all items are multiple-choice single-select items. 

	\item The test assesses students’ content and procedural knowledge across various content areas that are typical of physics education in middle schools. 

\end{romanlist}

Below, we report the study design we used for the development of our test. Subsequently, we report results, as well as selected parameters regarding the test’s psychometric quality.

\section{Study design}
For test development, we selected 40 physics-related items from the released TIMSS item sets for the primary- and middle-school levels.\cite{12,15,16,17} We translated items that were only published in English into German. Furthermore, through a discursive process, we adapted\footnote{Item adaptation and data collection for this study were provided by a pre-service physics teacher as part of his master’s thesis at the Universität Hamburg. For details, see Ref.~\refcite{18}. \label{fn1}} items with an open-ended format into multiple-choice single-select items (4–5 options) based on the items’ coding manual.\cite{15,16,17} 

The resulting pool of 40 items was administered\( ^{\mathrm{\ref{fn1}}} \) to N = 177 eighth grade students from Hamburg, Germany (mean age = 14.01 years; share of male participant = 52.54 \%). According to the TIMSS technical report\cite{14}, middle school students require about 1 minute to accomplish one multiple-choice single-select item, resulting in a total test time of approximately 40 minutes. In addition to their responses to these items, we also collected further background data from the participants. We surveyed their interest in science (see Ref.~\refcite{19}, p. 50; \( \alpha_{\mathrm{Cronbach}} \) = 0.87), their self-concept in science (see Ref.~\refcite{19}, p. 86; \( \alpha_{\mathrm{Cronbach}} \) = 0.84), their latest school grades in physics, and their cultural and economic capital (based on their number of books at home; see Ref.~\refcite{20}). Data collection was conducted from December 2018 to January 2019 at two middle schools in Hamburg and carried out in accordance with the legal and ethical standards for educational research at schools in the federal state of Hamburg.\cite{21} 

To evaluate our test’s psychometric quality, we conducted a Rasch analysis of the surveyed data\cite{22} using the R-package eRm version 1.0-2.\cite{23} We used the item-wise Wald test (split criterion: median) to check individual items for differential item functioning.\cite{24} In doing so, we detected poorly fitted items that needed to be removed from the test. After misfitting items were removed from the test, we applied Andersen’s likelihood ratio test (split criterion: median) to evaluate whether the Rasch model appropriately fits participants’ test-taking behavior\cite{24} and calculated the test’s item- and person-separation reliability.\cite{25} Finally, we conducted correlation analyses between the participants’ physics proficiency (estimated person parameters from the Rasch analysis) and their interest in science, their self-concept in science, their latest school grades in physics, and their cultural and economic capital in order to examine the correlational validity of our developed test.

\begin{sidewaystable*}
	\tbl{Summary of items included in the developed test.}
	{\begin{tabular}{@{}cccccccccccccc@{}}
			\toprule\\[-8pt]
			
			{} &  & & \multicolumn{6}{c}{response frequency (\%)} & & & & \\[3pt]
			{} &  & & \multicolumn{6}{c}{(correct response in {\bf bold})} & & & &  \\[3pt]
			\cline{4-9} \\[-6pt]
			
			item & origin & content area & A & B & C & D & E & \specialcell{no \\ response} & \specialcell{correct \\ response} & difficulty & \specialcell{infit \\ mean square} & \specialcell{outfit \\ mean square} & \specialcell{point-measure \\ correlation} \\[3.5pt]
			
			\hline\\[-6pt]
			
			1 & Ref.~\refcite{12}, p. 87 & thermodynamics & {\bf 89.27} & 3.95 & 3.39 & 3.39 & & 0.00 & A & -2.23 & 0.94 & 0.91 & 0.31 \\[7pt]
			2 & Ref.~\refcite{12}, p. 95 & optics & 19.77 &{\bf 50.28} & 17.51 & 1.13 & 3.95 & 7.34 & B & 0.04 & 1.01 & 0.94 & 0.45 \\[7pt]
			3 & Ref.~\refcite{12}, p. 89 & thermodynamics & 9.60 & 12.99 &{\bf 71.75} & 5.08 & & 0.56 & C & -0.91 & 1.12 & 1.13 & 0.31 \\[7pt]
			4 & Ref.~\refcite{12}, p. 94 & thermodynamics & 11.30 & 13.56 & 22.60 & {\bf 50.28} & & 2.26 & D & 0.16 & 0.99 & 0.97 & 0.46 \\[7pt]
			5 & Ref.~\refcite{12}, p. 90 & electrics & 10.73 &	12.99 &	{\bf 62.71} & 4.52 & 6.78 &	2.26 & C &	-0.47  &	0.93 &	0.86 &	0.49 \\[7pt]
			6 & Ref.~\refcite{15}, p. 45 & optics & {\bf 72.88} &	4.52 &	11.86 &	9.04 & & 1.69 & A & -1.02 & 0.92 & 0.98 & 0.45 \\[7pt]
			7 & Ref.~\refcite{12}, p. 92 & mechanics & 5.65 & 12.43 & 12.43 & {\bf 68.36} & & 1.13 & D & -0.72 & 0.83 & 0.82 &0.53 \\[7pt]
			8 & Ref.~\refcite{12}, p. 93 & optics & {\bf 73.45} & 9.04 & 10.73 & 4.52	 &  & 2.26 & A & -1.09 & 0.87 & 0.83 & 0.48 \\[7pt]
			9 & Ref.~\refcite{12}, p. 94 & acoustics & 7.34 & 12.99 & 10.73 & {\bf 64.97}  &  & 3.95 & D & -0.63 & 0.90 & 0.80 & 0.50 \\[7pt]
			10 & Ref.~\refcite{12}, p. 96 & mechanics & {\bf 27.12} & 19.21 & 29.38 & 22.03 &  & 2.26 & A & 1.38 & 1.16 & 1.32 & 0.28 \\[7pt]
			11 & Ref.~\refcite{12}, p. 102 & mechanics & 9.60 & 37.29 & 18.08 & {\bf 31.64} & & 3.39 & D & 1.12 & 1.02 & 1.14 & 0.41 \\[7pt]
			12 & Ref.~\refcite{12}, p. 99 & magnetism & {\bf 40.11} & 14.12 & 32.77 & 10.17 &  & 2.82 & A & 0.65 & 1.06 & 1.14 & 0.40 \\[7pt]
			13 & Ref.~\refcite{12}, p. 100 & mechanics & {\bf 18.64} & 30.51 & 24.29 & 24.29 &  & 2.26 & A & 1.97 & 0.98 & 1.16 & 0.36 \\[7pt]
			14 & Ref.~\refcite{12}, p. 101 & electrics & 10.17 & 8.47 & 23.16 & {\bf 46.33} & 5.65 & 6.21 & D & 0.30 & 0.84 & 0.77 & 0.58 \\[7pt]
			15 & Ref.~\refcite{15}, p. 40 & optics & {\bf 69.49} & 9.04 & 14.12 & 4.52 &  & 2.82 & A & -0.81 & 0.94 & 0.86 & 0.46 \\[7pt]
			16 & Ref.~\refcite{16}, p. 83 & mechanics & 14.69 & 14.69 & {\bf 48.02} & 14.12 & 3.95 & 4.52 & C & 0.21 & 1.03 & 1.09 & 0.42 \\[7pt]
			17 & Ref.~\refcite{16}, p. 47 & acoustics & 14.12 & 11.86 & 12.43 & {\bf 51.98} & 6.78 & 2.82 & D & 0.09 & 0.91 & 0.83 & 0.52 \\[7pt]
			18 & Ref.~\refcite{16}, p. 26 & magnetism & 9.60 & 25.42 & {\bf 41.81} & 16.95 &  & 6.21 & C & 0.54 & 1.08 & 1.08 & 0.39 \\[7pt]
			19 & Ref.~\refcite{15}, p. 49 & mechanics & 7.91 & 21.47 & 23.73 & {\bf 32.77} & 4.52 & 9.60 & D & 0.92 & 0.98 & 0.98 & 0.45 \\[7pt]
			20 & Ref.~\refcite{12}, p. 91 & optics & 10.17 & 18.08 & {\bf 42.94} & 22.60 &  & 6.21 & C & 0.49 & 1.04 & 1.13 & 0.41 \\[3pt]
			\Hline
	\end{tabular} \label{tbl1} }
\end{sidewaystable*}

\section{Results}
In total, 20 items were excluded from the test due to differential item functioning (detected via the item-wise Wald test). As summarized in Table \ref{tbl1}, the remaining 20 items do not show irregularities regarding their model fit indices. Both the infit and the outfit mean square of the items lie within a model-fitting range.\cite{22} In addition, all 20 items show a positive and sufficient point-measure correlation.\cite{26} Furthermore, the Andersen’s likelihood ratio test revealed that the Rasch model adequately captures participants’ responses (LR-value = 11.82, df = 19, p = 0.89). 

The remaining 20 items address different content areas typical of physics education in middle schools (electrics, magnetism, mechanics, optics, thermodynamics), with mechanics (six items) and \linebreak

\begin{figurehere}
	\centerline{
		\includegraphics[width=0.5\textwidth]{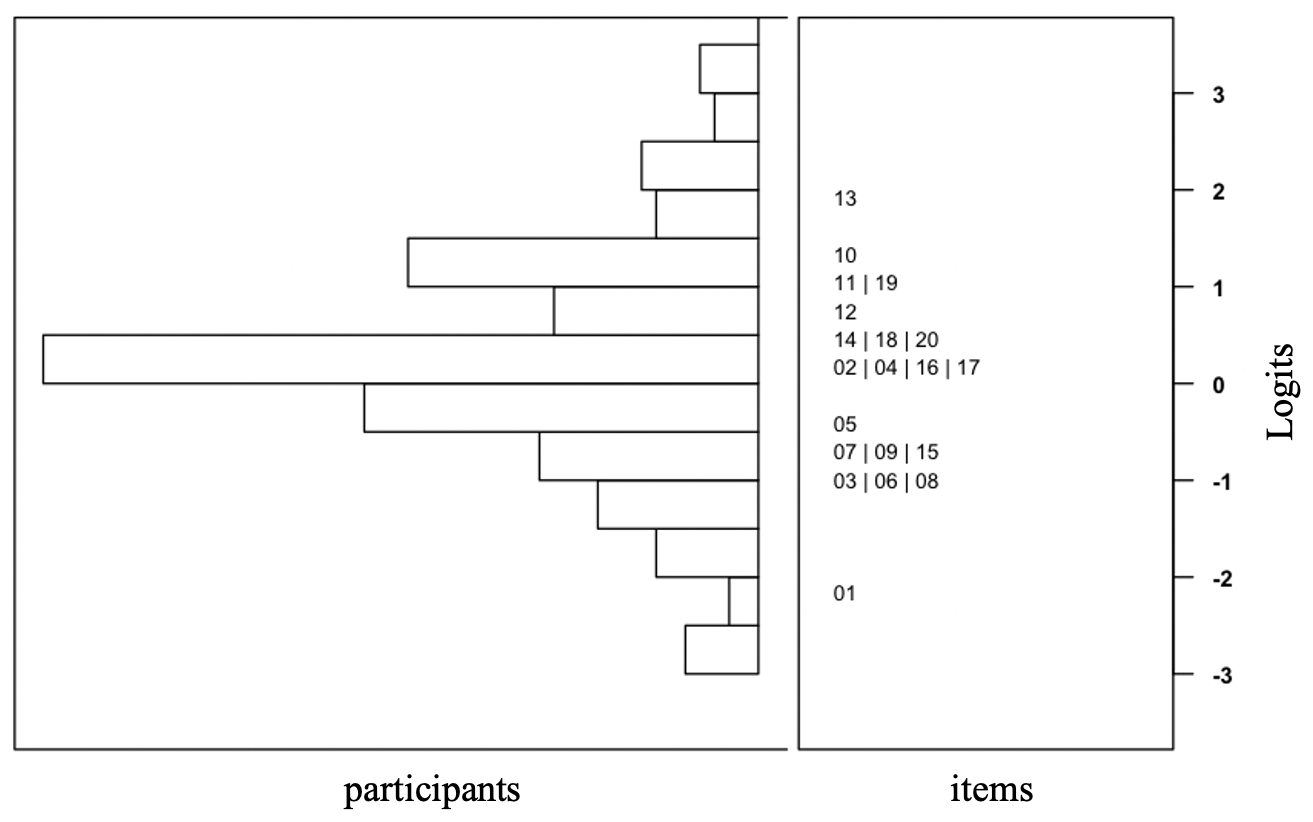}
	}
	\caption{Distribution of estimated item difficulties versus person parameters (Wright Map).}
	\label{fig1}
\end{figurehere}

\noindent optics (five items) being particularly prevalent (see Table \ref{tbl1}). Analysis of the Wright Map (Figure \ref{fig1}) indicates that all 20 items properly cover the range of the participants’ physics proficiency, especially within the proficiency mid-range. Only the lowest and uppermost ends of the Wright Map show noteworthy gaps between the distribution of estimated item difficulties versus person parameters. This result is also reflected in solid reliability indicators for our test. The person-separation reliability reached a coefficient of 0.78 and is thus sufficient.\cite{27} For the item-separation reliability, a coefficient of 0.97 was obtained. Therefore, our test’s item separation reliability can be considered excellent.\cite{27}

Regarding correlations with additional participant attributes, our results were in line with expectations in terms of our test’s correlative validity. We found a moderate and positive correlation between the participants’ estimated person parameters and their interest in science (\( r_{\mathrm{Pearson}} \) = 0.45; p \(<\) 0.01), as well as their self-concept in science (\( r_{\mathrm{Pearson}} \) = 0.34; p \(<\) 0.01). The rank correlation between estimated person parameters and latest school grades in physics was negative, as expected\footnote{The negative correlation between estimated person parameters and latest school grades in physics results from the German school grade scale that ranges from 1 = very good to 6 = insufficient.}, but low (\( \rho_{\mathrm{Spearman}} \) = -0.24; p \(<\) 0.01). For the participants’ cultural and economic capital, the rank correlation was also low but positive (\( \rho_{\mathrm{Spearman}} \) = 0.22; p \(<\) 0.01).

\section{Summary and Discussion}
The test developed in this study allows us to assess students’ physics proficiency at the middle-school level based on 20 translated (German) and adapted TIMSS items. Based on the TIMSS technical report\cite{14}, the test requires approximately 20 minutes to complete (1 minute per item), which is in accordance with our experiences. One particular advantage of the test is the multiple-choice-single-select format of all items because this format enables us to assess students’ physics proficiency in a highly test-economic and objective manner. 

Administering the developed test to N = 177 middle school students from Hamburg (Germany) and performing a Rasch analysis of the surveyed data revealed a sufficient person-separation reliability, an excellent item-separation reliability, and a satisfactory alignment between the distributions of estimated item difficulties and person parameters. These findings, as well as consistent results from our correlational analyses, provide cogent evidence that our developed test allows for a valid assessment of middle school students’ physics proficiency.

Finally, it should be noted that there are several limitations regarding the results of our study. We administered our developed test only to eighth grade students from Hamburg, Germany. Therefore, based on our results, it is not yet possible to draw valid conclusions regarding whether our developed test is suitable for students of higher or lower grades or within school systems significantly differing from the Hamburg education system. Furthermore, because electrics, which is a key topic within the physics curriculum for middle schools in Germany, is underrepresented among the test’s items (2 out of 20 items), it is reasonable to assume that our developed test captures students’ proficiency within this content area only to a limited extent. Consequently, in future research, it might be reasonable to extend our developed test by adding further items on electrics. Finally, as our Wright Map analysis revealed, it could be advantageous to extend our test by adding some items with a very high and/or very low item difficulty. Presumably, this could further improve the psychometric quality of our developed tests, particularly for assessing students with very high and/or low physics proficiency.

\section*{Note}
For non-commercial purposes, our developed test can be made available upon request by contacting the authors. 

\section*{Acknowledgments}
We thank the International Association for the Evaluation of Educational Achievement (IEA) for permitting us to reuse and revise TIMSS items within this study. 

\section*{Funding}
The research presented in this paper was supported by the VAMPS project, funded by the German Research Foundation (DFG).

\end{multicols}

\noindent\rule[1ex]{\textwidth}{1pt}

\paragraph{Markus Sebastian Feser} works as a postdoc at the Department of Education at the Universität Hamburg in Germany. His main research focus is in the professional development of (pre-service) science teachers. Among other topics, his current research focuses on the sense of belonging of students and student teachers to science, students’ conceptions of the viscous behavior of fluids, and the role of language in teaching and learning physics.

\paragraph{Dietmar Höttecke} works as a professor for physics education at the Department of Education at the Universität Hamburg in Germany. His main research focus is fostering students’ understanding the nature of science and the role of language in physics teaching and learning. He has led numerous research and development projects and authored textbooks for physics teacher education.

\end{document}